\documentclass[twocolumn,10pt,prl]{revtex4-1}
\usepackage[dvips]{graphicx}
\usepackage{amsmath,amsfonts,amssymb,bm,mathrsfs,graphicx,rotating,feynmp,pstricks,mathtools,cancel}
\usepackage{multirow}
\usepackage{verbatim}
\usepackage{hyperref}
\usepackage{color}
\usepackage{enumerate}
\usepackage[T1]{fontenc} % if needed
\usepackage[normalem]{ulem}
\usepackage{footnote}
\usepackage[caption=false]{subfig}

\usepackage{afterpage}

\newcommand{\beq}{\begin{equation}}
\newcommand{\eeq}{\end{equation}}

\usepackage{ifpdf}
\ifpdf
\fi

\renewcommand{\subsubsection}[1]{\addtocounter{subsubsection}{1}
\par\nobreak
\medskip
\nobreak
\noindent{\it \thesubsubsection.  #1 }
\par\nobreak\medskip\nobreak}

\def\lpar#1#2#3#4{\rlap{\raise#3\hbox{$\hskip#4#1\left\{\mbox{\phantom{\rule[0mm]{0mm}{#2}}}\right.$}}}
\def\rpar#1#2#3#4{\rlap{\raise#3\hbox{$\hskip#4\left\}#1\mbox{\phantom{\rule[0mm]{0mm}{#2}}}\right.$}}}

\begin{document}

\title{Searching for Displaced Higgs Decays}

\author{Csaba Cs\'aki${}^1$}
\email{csaki@cornell.edu}
\author{Eric Kuflik${}^1$}
\email{kuflik@cornell.edu}
\author{Salvator Lombardo${}^1$}
\email{sdl88@cornell.edu}
\author{Oren Slone${}^2$}
\email{shtangas@gmail.com}
\affiliation{${}^1$Department of Physics, LEPP, Cornell University, Ithaca, NY 14853, USA}
\affiliation{${}^2$Raymond and Beverly Sackler School of Physics and Astronomy, Tel-Aviv University, Tel-Aviv
69978, Israel
}

\begin{abstract}
We study a simplified model of the SM Higgs boson decaying to a degenerate pair of scalars which travel a macroscopic distance before decaying to SM particles. This is the leading signal for many well-motivated solutions to the hierarchy problem that do not propose additional light colored particles. Bounds for displaced Higgs decays below $10$~cm are found by recasting existing tracker searches from Run I. New tracker search strategies, sensitive to the characteristics of these models and similar decays, are proposed with sensitivities projected for Run II at $\sqrt{s} = 13 $ TeV. With 20 fb$^{-1}$ of data, we find that Higgs branching ratios down to $7 \times 10^{-4}$ can be probed for centimeter decay lengths.
\end{abstract}

\maketitle

\section{Introduction\label{sec:Intro}}
Run I of the LHC has severely constrained the parameter space for colored top partners that appear in traditional models which solve the hierarchy problem between the weak and Planck scales (theories such as supersymmetry or composite Higgs models). This has lead both theoretical and experimental studies to consider models that explain the hierarchy without additional light colored or SM-charged states. These are often referred to as models of ``uncolored''  or ``neutral'' naturalness, respectively.  The foremost example is the Twin Higgs~\cite{Chacko:2005pe,Barbieri:2005ri,Chacko:2005vw}, which provides a natural explanation by positing the existence of a dark sector containing colorless twin top partners, which partially cancel the quadratic divergence of the corrections to the Higgs mass. Other models of ``uncolored naturalness''  include folded supersymmetry and quirky little Higgs~\cite{foldedsusy,quirky}  where the top partners are color neutral but carry electroweak charge, while more general twin Higgs models go under the name of Orbifold Higgs~\cite{Craig:2014aea,orbifold}).

Twin Higgs models, as well as other models featuring neutral naturalness, generically predict the Higgs boson to decay to non-standard final state particles due to mixing between the ordinary and the twin Higgs. These particles may then either decay back to SM particles or leave the detector as missing energy. Such mixing is an essential ingredient for solving the hierarchy problem, which also results in corrections to the branching fractions of the already observed final states. However, the couplings of the twin states to the ordinary Higgs boson are expected to be quite small, making it difficult to test the presence of the twin sector by looking for deviations from Standard Model (SM) predicted rates~\cite{Carmi:2012in,Carmi:2012yp,Burdman:2014zta}. Thus, direct searches for Higgs decays leave this class of models largely unconstrained.

However, in many cases the twin matter from the Higgs decays traverses a macroscopic distance  before it decays back to SM matter, resulting in displaced Higgs decays~\cite{Juknevich:2009gg,Strassler:2006ri,fraternal,Kang:2008ea,Juknevich:2009ji,Curtin:2015fna}. The prediction of displaced vertices (DV) in these theories gives hope that such models may be discovered or constrained by searches in Run II of the LHC. The striking signature of displaced decays combined with negligible SM background allowed Run I LHC searches to place strong constraints on many beyond the SM signals which contain long-lived particles~\cite{Aad:2015rba,Aad:2015uaa,Aad:2015asa,CMS:2014wda}. Searches have been performed by ATLAS and CMS for decays in the tracker, calorimeter, and muon spectrometer. However, the benchmark models for the tracker searches have typically involved the production of heavy particles with $m \gtrsim 200$ GeV, possibly decaying to lighter particles with $m \gtrsim 50$ GeV. Large $p_T$ trigger thresholds and vertex requirements make these searches very inefficient for a 125 GeV Higgs boson decaying to hidden sector particles with masses $\lesssim 60$ GeV. As a result, exotic displaced decays of the Higgs boson within the ATLAS or CMS inner detector are weakly constrained, especially for the case of light intermediate particles. Furthermore, existing searches for decays in the calorimeter or muon spectrometer, which are sensitive to long-lived light particles, only constrain signals which produce at least two displaced decays per event, a feature which is not generic to these models. Depending on the details of the hidden sector, the signal may prefer only one metastable particle per event.

For Run II, displaced decays will continue to be an important signature for discovery of new physics. The aim of this paper is to present bounds for displaced Higgs decays by recasting existing tracker searches from Run I, and more importantly, to present projected sensitivity for Run II with $\sqrt{s} = 13$ TeV and propose new search strategies which are sensitive to the characteristics of these models.

Dedicated displaced triggers for Run II are efficient for detecting events with displaced Higgs decays for lifetimes ranging from $1 \text{ cm}$ to $1 \text{ m}$, while lepton triggers or vector boson fusion (VBF) triggers, which do not require displaced tracks, can be sensitive to shorter lifetimes. We project bounds both by using existing search techniques and by proposing new tracker search strategies using displaced jets together with jet substructure as a general method for detecting decays from light, long-lived particles.
In this work, we find Run II inner detector searches can probe such signatures down to sub-percent Higgs branching ratios with an integrated luminosity of 20 fb$^{-1}$.

\section{Models and their Signal}
\begin{figure}[t!]
\begin{center}
\vspace{1cm}
\begin{fmffile}{feyngraph2}
  \begin{fmfgraph*}(50,25)
    \fmfleft{i1,i2}
    \fmfright{o1,o2,o3,o4}
    \fmf{fermion}{i1,v4}    \fmflabel{$p$}{i1}
    \fmf{fermion}{i2,v4}    \fmflabel{$p$}{i2}
    \fmfblob{.16w}{v4}
    \fmf{dashes,lab=$H$}{v4,v1}
    \fmf{dots,label.side=right,label=$\pi_v$}{v1,v2}
    \fmf{dots,label.side=left,label=$\pi_v$}{v1,v3}
    \fmf{plain}{v2,o1}    \fmflabel{$\bar{f}$}{o1}
    \fmf{plain}{v2,o2}    \fmflabel{${f}$}{o2}
    \fmf{plain}{v3,o3}    \fmflabel{$\bar{f}$}{o3}
    \fmf{plain}{v3,o4}    \fmflabel{${f}$}{o4}

    \fmffreeze
   \renewcommand{\P}[3]{\fmfi{plain}{vpath(__#1,__#2) shifted (thick*(#3))}}
    \P{i1}{v4}{2.25,-.75}
    \P{i1}{v4}{-2,1}
    \P{i2}{v4}{2.4,.5}
    \P{i2}{v4}{-2.1,-1.35}
  %  \P{i2}{v4}{-2,1}

  \end{fmfgraph*}
\end{fmffile}
\vspace{.5cm}
\end{center}
\caption{ \label{fig:process}The Higgs boson, once produced, decays to two long-lived invisible scalars, $\pi_v$, which then decay back to visible particles via an off-shell Higgs.
}
\end{figure}
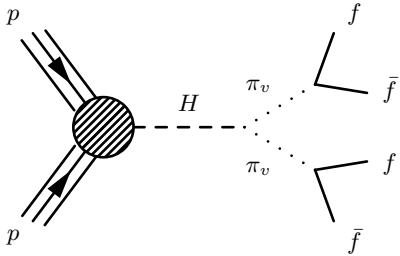

The phenomenology of Twin Higgs-type models is of the hidden valley type~\cite{Juknevich:2009gg,Strassler:2006ri,Juknevich:2009ji,Han:2007ae}, where twin particles communicate with the SM only via a Higgs portal, i.e. a mixing between the SM Higgs and its twin partner. This mixing can allow for production of twin particles following Higgs production at the LHC. If these particles are metastable on detector scales, they may traverse a macroscopic distance and decay back to SM particles within the detector, appearing as either a DV or as a decay in the hadronic calorimeter or muon spectrometer. The prototypical model we have in mind is that of the ``fraternal twin Higgs''~\cite{fraternal} where the twin QCD has no light quark generations, only a twin 3rd generation, resulting in metastable twin glueballs and twin bottomonia. Other models of neutral naturalness with top partners charged under the electroweak group will necessarily have similar phenomenology: the twin glueballs must be the lightest states of the hidden sector, in order to avoid direct constraints from LEP, resulting in the phenomenology considered here.
Typically, the mass of the intermediate scalars in the Higgs decays (corresponding to the twin glueballs or twin bottomonia) are expected to be in the $10-60$ GeV range~\cite{Curtin:2015fna}, depending on the details of the twin confining sector.

While the Twin Higgs is our motivation, we focus on a simplified signal of the SM Higgs boson decaying to a degenerate pair of hidden scalars, $\pi_v$, which travel a finite distance before decaying to SM particles (see Fig.~\ref{fig:process}). We assume that the decay occurs via mixing with the Higgs. Therefore, the couplings of the scalars, $\pi_v$, to SM particles are proportional to the Higgs' couplings. The dominant final states are $b\bar{b}$, $\tau^+ \tau^-$ and $c\bar{c}$, with the ratios 85:8:5 for $m_{\pi_v} \gtrsim 20$ GeV. 

We also present results for the scenario where the Higgs boson decays to a pair of degenerate hidden scalars, one of which is stable and escapes the detector. This signal could be realized if the branching ratio of the twin gluon to the metastable $0^{++}$ glueball is very small, in which case most twin glueballs produced will be stable since they do not have the have the right quantum numbers to mix with the SM Higgs.

\section{Bounds from Run I}
\begin{figure}[t!]
\center
\includegraphics[width=.47\textwidth]{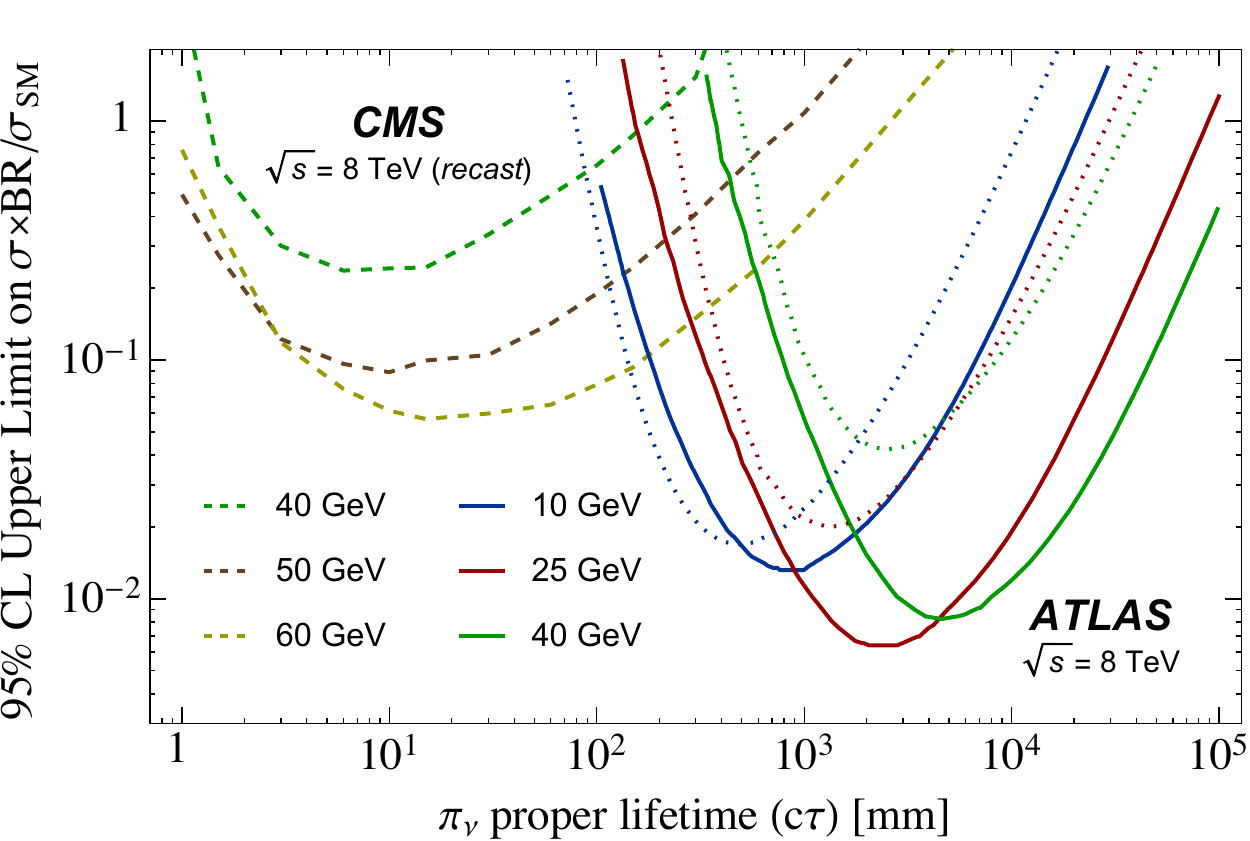}
\caption{95\% CL exclusion curves for Run I of the LHC. The ATLAS constraints are reproduced from the searches for long lived particles decaying in the muon spectrometer~\cite{Aad:2015uaa}  (solid) and the hadronic calorimeter~\cite{Aad:2015asa} (dotted). The CMS recast exlcusion curves are derived from the search displaced dijet in the inner tracker~\cite{CMS:2014wda} (dashed).} \label{8TeV}
\end{figure}

Several searches for displaced decays within the ATLAS and CMS detectors have been performed on Run I data.
In particular, two searches at  $\sqrt{s}=8$~TeV with the ATLAS detector have been interpreted for signals of the type studied here. The first of these is an ATLAS search~\cite{Aad:2015uaa} for two DVs either within the tracker or the muon spectrometer. This search is mostly sensitive to lifetimes of $\mathcal{O}(1\, {\rm m})$. The reason for this is that the trigger efficiencies are enhanced for DVs occurring in the muon spectrometer, while the trigger thresholds and strict vertex requirements applied for decays within the tracker are rarely satisfied by this signal. The second ATLAS search~\cite{Aad:2015asa} looks for low electromagnetic fraction jets indicative of decays within the hadronic calorimeter or at the edge of the electromagnetic calorimeter. This search is sensitive to similar lifetimes, but is not as powerful as the former search in constraining the branching ratio for all except the lightest intermediate scalar masses. This is due to a weaker trigger efficiency. These ATLAS exclusion curves are reproduced in Fig.~\ref{8TeV}.

The CMS search for displaced dijets~\cite{CMS:2014wda} and ATLAS multitrack DV searches~\cite{Aad:2015rba} look for decays within the tracker and constrain lifetimes ranging from $1 \text{ mm}$ to $1 \text{ m}$. We have fully recast these searches and interpreted the results in terms of displaced Higgs decays. We simulate the three largest production modes for the Higgs: gluon-gluon fusion (ggF), vector boson fusion (VBF), and vector associated production (VH). Hard processes are simulated in Madgraph~5~\cite{Alwall:2011uj} followed by hadronization and parton showering using Pythia~8~\cite{Sjostrand:2014zea}. We use Delphes~3~\cite{deFavereau:2013fsa} for the detector simulation with default efficiencies for the ATLAS and CMS detectors, excluding tracking efficiency. FastJet~\cite{Cacciari:2011ma} is used to cluster jets and apply jet substructure algorithms. Additional details of the simulations can be found in~\cite{Csaki:2015uza}.

For the ATLAS multitrack DV searches, we find that the trigger thresholds for leptons, jets, and MET are too strong for this search to be sensitive to a 125 GeV Higgs, including all Higgs production modes, and thus no further bound is established. The dilepton DV search performed in this study is efficient for the $\pi_v \rightarrow \mu^+ \mu^-$ decays but gives no bound due to the small expected branching ratio of this decay mode. However, the CMS displaced dijet search, while having low efficiencies, does have some sensitivity to displaced Higgs decays for the heavier range of the intermediate $\pi_v$ masses, for branching fractions as low as 5 percent in the $1 \text{ mm}$ - $1 \text{ m}$ proper lifetime regime. The reason for the somewhat increased sensitivity is that the CMS study takes advantage of a dedicated displaced trigger which allows for lower jet $p_T$ trigger thresholds by requiring two jets with $p_T > 60$ GeV to have displaced tracks with transverse impact parameter (IP) larger than 0.5 mm. The trigger is seeded by the level one requirement of scalar transverse energy, $H_T > 300$ GeV. However, this large $H_T$ requirement of the trigger preferentially selects events containing a boosted Higgs or large initial state radiation (ISR), which also results in boosted $\pi_v$'s, merging their decay products into a single jet. For these reasons, this search, which requires two jets associated to a DV, is not very efficient for the signal we consider. Furthermore, the vertex requirement, $m_{DV}$ $ > 4 $ GeV, and the background discriminant which prefers a large DV track multiplicity, decrease the efficiency for signals with light $\pi_v$. Nevertheless, we still find that the search places bounds on signals with $m_{\pi_v} \gtrsim 40$ GeV for lifetimes complementary to those obtained from the ATLAS searches.  Our resulting bounds on the Higgs branching fractions as a function of the $\pi_v$ lifetime obtained from our recast of the CMS dijet search (together with the previous ATLAS bounds) are presented in Fig.~\ref{8TeV}. We find limits for heavy $\pi_v$ and  shorter lifetimes, ranging from $1-1000 \text{ mm}$, that are somewhat weaker than the corresponding ATLAS bounds for longer lifetimes, while signals with $m_{\pi_v} \lesssim 40$ GeV remain unconstrained for lifetimes below $100 \text{ mm}$. We emphasize that these constraints do not apply for signals which only produce a single DV per event. For the case where one of the hidden particles is stable, the CMS dijet search does not have sensitivity since the events fail to pass the large $H_T$ requirement.

\section{Search Strategies and Projected Sensitivity for Run II}
In this section we propose new search strategies for detecting displaced Higgs decays within the ATLAS or CMS inner detector, noting that  lifetimes corresponding to the decay lengths considered here are mostly unconstrained. For longer decay lengths, a search for decays in the muon spectrometer would be more sensitive and the strategies considered here could be slightly altered and applied 
in order to achieve sensitivity to events with a single displaced decay.

There are major difficulties in detecting the signal under consideration due to the relatively light mass of the Higgs boson and of the hidden sector particles. To make matters worse, the dominant production mechanisms (ggF followed by VBF) tend to produce the Higgs boson close to rest. Therefore, a search with sensitivity to such a signal must either use a trigger with low $p_T$ requirements, possibly by taking advantage of a dedicated displaced trigger, or be restricted to boosted Higgs kinematics and pay the price of a relatively small production rate. Furthermore, DV searches, looking for single decays in the tracker, typically impose strict vertex requirements designed to cut out background events. However, signals with light intermediate particles often do not pass these requirements and cannot be detected by a generic DV search. For these reasons, model-specific searches are required in order to detect such signals.
Specifically, a successful search strategy should be designed with weak vertex requirements in order to enhance the number of expected signal events while retaining as low background as possible by imposing other event selection criteria.

An important point regarding the expected LHC phenomenology of Twin Higgs models is that the $\pi_v$ particles decay to the SM via a Higgs portal with final states which are expected to often be $b\bar{b}$. This has a few important consequences. First, one can search for decay products of the $b\bar{b}$ in conjunction with a DV, for example a muon or a dijet. It has been shown that requiring a muon within a cone of a displaced jet significantly reduces the displaced jet background~\cite{Abazov:2009ik, Aad:2013txa}. Requiring a displaced dijet associated to a DV was used as a background discriminant in the CMS displaced dijet search~\cite{CMS:2014wda}. Furthermore, depending on $m_{\pi_v}$, the displaced dijet can become merged into a single jet with many displaced tracks, resulting in an ``emerging jet'' signature~\cite{Schwaller:2015gea}. The merged jets typically exhibit a 2-prong substructure which can be used to reconstruct the displaced dijet, thus extending the displaced dijet search strategy to scenarios with light hidden sector particles. Some percent of events may contain two displaced vertices. This has been taken advantage of in searches performed by ATLAS~\cite{Aad:2015uaa,Aad:2015asa}. These signatures, together with a requirement of reconstructing the original Higgs mass and/or the $\pi_v$ masses in the displaced jets of the event, could be used to remove background events. In what follows, we present the details of the search strategies considered in this work.

\subsection{Triggers}
For Run I, triggers were a major limitation on placing constraints on $\mathcal{O}$(cm) lifetimes. This situation may significantly improve for Run II due to the possible implementation of improved dedicated displaced triggers. These allow for lower $p_T$ thresholds, giving better sensitivity to Higgs events with displaced decays. Triggers based on the production process of the Higgs, either VBF or VH, do not strongly depend on the lifetime of the $\pi_v$ particles and may be useful for probing all lifetimes, in particular short lifetimes not picked up by displaced triggers. The five triggers which have been considered in this study are detailed in Table~\ref{tab:trigger_requirements}.

CMS has several displaced triggers implemented for Run II, including a dedicated displaced trigger with thresholds designed to pick up Higgs events with displaced jets, which we will refer to as the ``displaced jet'' trigger.
This trigger requires a VBF signature, along with a displaced jet containing tracks with IP $>$ 2 mm.

\begin{table}
\begin{tabular}{| l |p{6cm}|}
\hline
Trigger & Trigger Requirement \\
\hline
Displaced jet~\footnotemark[1] & $H_T$ > 175 GeV or three jets with
$p_T^{j_{1,2,3}} > (92,76,64)$ GeV, $|\eta_{j_{1,2,3}}| < (5.2,5.2,2.6)$ with $|\eta_{j_1}|$ or $|\eta_{j_2}| < 2.6$, and two jets satisfying $m_{jj}$ > 500 GeV and $\Delta \eta$ > 3.0.
A displaced jet satisfying $p_T$ > 40 GeV, at most 1 prompt track (2D IP < 2.0 mm), and at least 2 displaced tracks. \\
\hline
Inclusive VBF & Two jets with $|\eta_{j_1,j_2}| > 2$, $\eta_{j_1}\cdot\eta_{j_2}<0$, $|\eta_{j_1} - \eta_{j_2}| > 3.6$ and $m_{j_1,j_2} > 1000$ GeV. \\
\hline
VBF, h $\rightarrow$ $b\bar{b}$ & Three jets with $p_T^{j_{1,2,3}} > (112,80,56)$ GeV and $|\eta_{j_{1,2,3}}| < (5.2,5.2,2.6)$ and at least one of the two first jets with $|\eta_{j_1}|$ or $|\eta_{j_2}| < 2.6$. \\
\hline
Isolated Lepton & One lepton with $p_T > 25$ GeV, $|\eta| < 2.4$, and 3D IP < 1 mm. Isolation requires the summed $p_T$ of all tracks with $p_T>1$ and within $\Delta R<0.2$ of the lepton is less than 10\% of the lepton $p_T$.\\
\hline
Trackless jets & A jet with $p_T > 40$ GeV and $|\eta| < 2.5$ matched with a muon with $p_T > 10$ GeV within $\Delta R = 0.4$.  No tracks with $p_T > 0.8$ GeV in the ID within a $\Delta \phi \times \Delta \eta$ region of $0.2 \times 0.2$. \\
\hline
\end{tabular}
\caption{ \label{tab:trigger_requirements} Triggers for Run II which may be sensitive to displaced Higgs decays.}
\footnotetext[1]{A previous version of this paper had incorrectly stated and used the trigger requirement 2D IP < 2.0 cm and did not apply the $m_{jj}$ requirement to events passing the $H_T$ requirement.}
\end{table}

\begin{table*}
\centering
\begin{tabular}{|l|c|c|c|c||c||c|c|c||c||c|c|c||c||}
\hline
\multicolumn{1}{|c|}{\multirow{2}{*}{\textbf{Trigger}}} & \multicolumn{1}{|c|}{\multirow{2}{*}{$\mathbf{m_{\pi_v}}$ \textbf{(GeV)}}} & \multicolumn{4}{|c|}{$\mathbf{c\tau = 1 \textbf{ mm}}$} & \multicolumn{4}{|c|}{$\mathbf{c\tau = 10 \textbf{ mm}}$} & \multicolumn{4}{|c|}{$\mathbf{c\tau = 100 \textbf{ mm}}$} \\
\cline{3-14}
 & & $\mathbf{\epsilon_{ggF}}$ & $\mathbf{\epsilon_{VBF}}$ & $\mathbf{\epsilon_{VH}}$ & $\mathbf{\epsilon_{Total}}$ & $\mathbf{\epsilon_{ggF}}$ & $\mathbf{\epsilon_{VBF}}$ & $\mathbf{\epsilon_{VH}}$ & $\mathbf{\epsilon_{Total}}$ & $\mathbf{\epsilon_{ggF}}$ & $\mathbf{\epsilon_{VBF}}$ & $\mathbf{\epsilon_{VH}}$ & $\mathbf{\epsilon_{Total}}$ \\
\hline
\multirow{3}{*}{Displaced jet} & 10 & $0.03\%$ & $1.3\%$ & $1.1\%$ & \textbf{0.2\%} & 1.0 \% & 30.0\% & 25.1\% & \textbf{3.9\%} & 1.0\% & 42.0\% & 34.7\% & \textbf{5.1\%} \\
& 25 & 0.01\% & 0.8\%& 0.7\% & \textbf{0.09\%} & 0.7\% & 20.4\% & 16.9\% & \textbf{2.7\%} & 1.5\% & 45.3\% & 37.3\%& \textbf{5.9\%} \\
& 40 & 0.02\% & 1.0 \% & 0.9\% & \textbf{0.1\%} & 0.6\% & 19.7\% & 16.4\% & \textbf{2.5\%} & 1.4\% & 44.6\% & 36.3\% & \textbf{5.7\%} \\
\hline
\multirow{3}{*}{Inclusive VBF} & 10 & $1.9\%$ & $15.5\%$ & $0.8\%$ & \textbf{2.8\%} & $1.8\%$ & 15.5\% & $0.7\%$ & \textbf{2.8\%} & 1.6\% & 15.1\% & 0.6\% & \textbf{2.6\%} \\
& 25 & 1.7\% & 15.3\% & 0.7\% & \textbf{2.7\%} & $1.7\%$ & $15.3\%$ & $0.7\%$ & \textbf{2.7\%} & 1.6\% & 15.2\% & 0.6\% & \textbf{2.6\%} \\
& 40 & 1.6\% & 15.2\% & 0.7\% & \textbf{2.6\%} & 1.6\% & 15.2\% & 0.7\% & \textbf{2.6\%} & 1.6\% & 15.2\% & 0.6\% & \textbf{2.6\%} \\
\hline
\multirow{3}{*}{VBF, h $\rightarrow$ $b\bar{b}$} & 10 & $5.8\%$ & 20.3\% & $13.1\%$ & \textbf{7.2\%} & 5.8\% & 20.2\% & 13.0\% & \textbf{7.2\%} & 3.5\% & 13.3\% & 8.1\% & \textbf{4.4\%} \\
& 25 & $4.6\%$ & $16.6\%$ & 10.9\% & \textbf{5.8\%} & $4.7\%$ & 16.7\% & 10.9\% & \textbf{5.9\%} & 4.2\% & 15.2\% & 9.7\% & \textbf{5.3\%} \\
& 40 & 4.0\% & 14.2\% & 9.2\% & \textbf{5.0\%} & 4.0\% & 14.2\% & 9.2\% & \textbf{5.0\%} & 3.8\% & 13.9\% & 8.9\% & \textbf{4.8\%} \\
\hline
\multirow{3}{*}{Isolated Lepton} & 10 & 3.6\% & 3.7\% & 14.7\% & \textbf{4.1\%} & 1.0\% & 1.0\% & 12.5\% & \textbf{1.5\%} & 0.1\% & 0.2\% & 11.8\% & \textbf{0.6\%} \\
& 25 & 1.0\% & 1.5\% & 13.0\% & \textbf{1.6\%} & 0.3\% & 0.4\% & 11.9\% & \textbf{0.8\%} & 0.05\% & 0.07\% & 11.7\% & \textbf{0.6\%} \\
& 40 & 1.0\% & 1.4\% & 12.6\% & \textbf{1.6\%} & 0.3\% & 0.4\% & 11.9\% & \textbf{0.8\%} & 0.05\% & 0.07\% & 11.6\% & \textbf{0.6\%} \\
\hline
\multirow{3}{*}{Trackless jet} & 10 & 0.02\% & 0.04\% & 0.04\% & \textbf{0.02\%} & 0.8\% & 1.5\% & 1.3\% & \textbf{0.9\%} & 2.0\% & 2.4\% & 2.2\% & \textbf{2.0\%} \\
& 25 & 0.02\% & 0.04\% & 0.06\% & \textbf{0.02\%} & 0.5\% & 1.0\% & 0.8\% & \textbf{0.6\%} & 3.6\% & 5.9\% & 5.0\% & \textbf{3.8\%} \\
& 40 & 0.01\% & 0.02\% & 0.03\% & \textbf{0.01\%} & 0.1\% & 0.2\% & 0.2\% & \textbf{0.1\%} & 2.1\% & 4.1\% & 3.3\% & \textbf{2.3\%} \\
\hline
\end{tabular}
\caption{ \label{tab:triggers}A comparison of trigger acceptances for $m_{\pi_v} = 10,\ 25,\ 40$ GeV and $c\tau = 1,\ 10,\ 100$ mm. The acceptance is given for Higgs production via ggF, VBF and VH. The rightmost column for each lifetime is the total acceptance, $\epsilon_{Total}$, weighted by the cross sections for the various production mechanisms.}
\end{table*}

The full trigger requirements for this ``displaced jet'' trigger are detailed in Table~\ref{tab:trigger_requirements}.
The trigger is efficient for all production processes of the Higgs, provided $c \tau_{\pi_v} \gtrsim 2$ mm.
For short lifetimes, this trigger's efficiency drops since the displacement requirement is often no longer satisfied. CMS has other dedicated displaced triggers on their Run II trigger menu. The trigger with second lowest thresholds requires two displaced jets with tracks satisfying IP > 0.5 mm (analogous to the Run I triggers used in~\cite{CMS:2014wda}) at the cost of higher $H_T$ and jet $p_T$ requirements. A search with this trigger could be sensitive to shorter lifetimes, however the search would be restricted to events with boosted Higgs events, rendering it inefficient for the signal considered in this study.

For short lifetimes of $\mathcal{O}(10 \text{ mm})$, non-displaced triggers are more efficient. VBF or lepton triggers may be useful for a DV search in this scenario. One possibility is to search for a VBF signature, picking out two jets with large invariant mass. This has been denoted as the ``Inclusive VBF'' trigger in Table~\ref{tab:trigger_requirements}. Another possibility is a trigger designed to pick up VBF production with the Higgs decaying into a $b\bar{b}$ pair. Such a trigger, which requires three jets with varying $p_T$ and $\eta$, is denoted as the ``VBF, h$\rightarrow b\bar{b}$'' trigger in Table~\ref{tab:trigger_requirements}. Yet another possibility is to trigger on prompt leptons from the leptonic decay of W or Z in VH events or on displaced leptons from the $b\bar{b}$ meson decays. Such a trigger, with a low lepton $p_T$ threshold, would likely require isolation cuts under Run II conditions, giving a significant reduction in ggF and VBF efficiency for which the leptons typically arise from $b\bar{b}$ decays and are not isolated. As a conservative estimate for such a trigger we impose track isolation cuts from \cite{ATLAS:2012yna} and impose a cut on the 3D IP of the lepton as detailed in Table~\ref{tab:trigger_requirements}.

ATLAS has implemented a trackless jets trigger~\cite{Aad:2013txa} which may be used to pick up decays occurring beyond the pixel layers of the tracker (the outermost layer is located at $r=10.2$ cm). This trigger is denoted the ``trackless jet'' trigger in Table~\ref{tab:trigger_requirements}. Tracks originating beyond the pixel layer are not reconstructed by the level two trigger. Decays occurring outside of the pixel layers thus give a signature of a jet which is isolated from reconstructed tracks. We find that this trigger is inefficient for reconstructing displaced vertices within the tracker. The trigger has small efficiency due to the requirement of a muon matched to the jet. Also, the trigger preferentially selects events in which the displaced decay is outside the pixel layers where track reconstruction efficiencies are lower, making vertex reconstruction more difficult. For search strategies which require a muon matched to a displaced jet, this trigger is more competitive with a single dedicated displaced jet trigger, but again preferentially selects longer lifetimes.

By triggering on the production process and not requiring displaced tracks at the trigger level, the VBF and lepton triggers are sensitive to all lifetimes. However, the displaced jet trigger may have significantly less background than a pure VBF trigger due to the displaced requirement. In this study, we present results for each of two such triggers separately. We utilize the ``displaced jet'' trigger and the ``VBF h$\rightarrow b\bar{b}$'' trigger. We find this combination provides the best sensitivity over a wide range of lifetimes, however background considerations may lead to the use of alternative triggers.

The acceptance of each of the triggers described above is given in Table~\ref{tab:triggers} for signal events which follow Higgs production via ggF, VBF and VH.  For jet clustering, we use the anti-k$_t$ algorithm with $\Delta R = 0.5$.
The vertex reconstruction efficiency is expected to be approximately independent of the trigger acceptance for a given decay length and choice of vertex selection cuts. However, the displaced jet and trackless jet triggers preferentially accept events with longer decay lengths which have lower vertex reconstruction efficiency.
It may be possible to reduce vertex requirements as a result of using a lower background trigger, $\textit{e.g.}$ a lepton trigger together with reconstructing the W/Z boson in VH events is known to cut down on multijet background~\cite{Kaplan:2011vf}.

\subsection{Event Reconstruction}
\begin{figure*}[t!]
\center
\includegraphics[width=.48\textwidth]{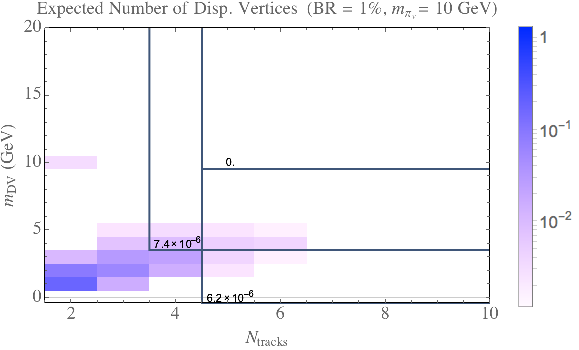}
\hfill
\includegraphics[width=.48\textwidth]{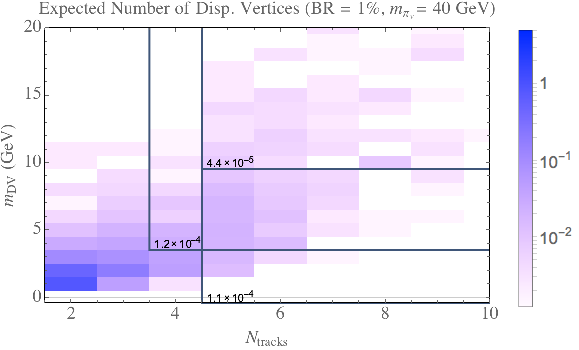}
\caption{Expected number of signal displaced vertices as a function of vertex track multiplicity and invariant mass in events passing the ``VBF, h$\rightarrow b\bar{b}$'' trigger at $\sqrt{s} = 13$ TeV, with an integrated luminosity of 20 fb$^{-1}$, and assuming a Higgs branching ratio of 1\%. The two sample signals correspond to $m_{\pi_V} = 10$ GeV (left), 40 GeV (right) with $c \tau = 100$ mm. The signal regions considered in this study are boxed and labeled with an overall efficiency for an event to satisfy the trigger requirement and contain at least one vertex in the signal region.} \label{fig:DVselection_10GeV}
\end{figure*}

Provided events pass the triggers, the next difficulty in detecting displaced $\pi_v$ decays is to reconstruct the DV and apply event selection requirements to remove background events. The details of DV reconstruction and selection will be very important for signals with long-lived particles with masses below $\sim 20$ GeV for which tracker vertex reconstruction efficiencies depend strongly on $m_{\pi_v}$.
Due to the light mass of the $\pi_v$ particles, the signal vertices have characteristically low invariant mass, $m_{DV}$,  and track multiplicity, $N_{tracks}$, even with perfect tracking efficiency. Combining the low track multiplicity of the signal with reduced tracking efficiency for displaced tracks makes displaced vertex reconstruction within the tracker difficult.
For example, with $m_{\pi_v} \lesssim$ 20 GeV, a 5-track, $m_{DV} > 10$ GeV single multitrack DV search would have very limited sensitivity. A search which requires two reconstructed 5-track DVs within the tracker would also have weak sensitivity due to the low probability to reconstruct multiple 5 (or more) track vertices in the same event. 
A primary goal of this paper is to weaken the vertex mass and track requirements by requiring additional objects associated to the DV or other event selection criteria in order to achieve sensitivity to light signals. If these selection criteria are not able to reduce backgrounds to zero, it may be necessary to search for the signal as an excess in background events.

Background vertices can arise from heavy flavor decays, interactions with material in the detector, and accidental crossing of tracks. Background DVs typically have small invariant mass and low track multiplicity~\footnote{See for instance, figures 9-10 in \cite{Aad:2015rba}.}. Heavy flavor decays have low $m_{DV}$ and are associated to tracks with small impact parameters  (IP $\lesssim$ 0.5 mm). Requiring displaced tracks significantly suppresses SM backgrounds. However, for sub-mm lifetimes and for searches with low $m_{DV}$ requirements, SM backgrounds can be important. We follow the procedure in~\cite{Aad:2015rba} and only vertex tracks which are significantly displaced with IP > 2 mm and impose $m_{DV}$ requirements such that SM backgrounds are not the dominant source of background. Detector material interactions can be reduced by mapping out regions of detector material as was done by ATLAS \cite{Aad:2015rba} and removing vertices which are within or close to regions of dense material. It will be crucial for CMS to perform such a detector study in order to be sensitive to light $\pi_v$ signals. In our projected searches, we remove DVs which occur in material areas of the detector using the ATLAS maps from \cite{Aad:2015rba}. Since the overall signal efficiency is not very sensitive to the exact details of the material region, the same map has been used for the purpose of projecting searches for the CMS detector.

As mentioned above, standard selection criteria designed to remove background DVs also remove most, or in some cases all of the signal considered in this paper. Given the incomplete detector simulations performed in this study, we are not able to simulate background events which are dominated by the full detector response and details of the track and vertex reconstruction procedures. We try to retain vertex requirements similar to those in existing searches and assume one background event as a benchmark for projecting sensitivities. This assumption may not be valid for new triggers and search strategies with weaker selection requirements. However, as mentioned above, it may be possible to search for an excess in background events, in which case the results would be weakened according to the number of background events expected.

The signal has several features which can be used as a background discriminant. Two long-lived $\pi_v$ particles are produced in each event. The search for DVs in the tracker and MS by ATLAS \cite{Aad:2015uaa} (their tracker search did not have appropriate triggers for this signal) and tracker searches performed by CDF~\cite{Aaltonen:2011rja} and D0~\cite{Abazov:2009ik} avoided an $m_{DV}$ requirement by selecting events with two vertices. While this strategy is sensitive to the main signal considered in this paper, we also consider strategies which require only one DV per event in order to retain sensitivity to a more general class of models, some of which predict one DV together with MET or additional hadronic jets.
Additionally, the $\pi_v$ decay mostly to $b\bar{b}$. Searches can take advantage of this fact by looking for a dijet emerging from a DV or muons from the $b$ decays pointing back to the DV.

Most of the displaced decays are to $B$-mesons which are then further displaced from the vertex of the $\pi_v$ decay causing a degradation in efficiency in cases where the two $B$-meson vertices reconstruct as separate vertices. This can occur in central parts of the tracker with higher vertex resolution and could be used to detect the signal by searching for two nearby DVs, perhaps in the same displaced jet if the $\pi_v$ is boosted. We do not explore this option since we cannot estimate the displaced vertex resolution without the full detector simulation and procedure. A search of this kind, looking for emerging jets with multiple DVs within the same jet, has not been performed yet.

In order to maximize the signal efficiency for light $\pi_v$ signals, it is important to merge nearby vertices for the scenario where both $B$-meson decays reconstruct as separate, nearby vertices. Merging nearby vertices introduces a new source of background: two nearby background vertices with low track multiplicity can become merged and pass the selection requirements.
Motivated by {ATLAS}~\cite{Aad:2015rba}, we merge all vertices within 1 mm of each other and merge two vertices with $N_{tracks} \geq 3$ and $N_{tracks} \geq 2$, respectively, if the vertices are within 5 mm of each other in order to retain sensitivity to events where the $\pi_v$ decay reconstructs as two separate vertices.

In Fig.~\ref{fig:DVselection_10GeV} we show the number of expected DVs for a signal sample of $m_{\pi_v} = 10$ and $40$ GeV both with $c\tau = 100$ mm for a 20 fb$^{-1}$ dataset, as a function of $N_{tracks}$ and $m_{DV}$, passing the ``VBF, h$\rightarrow b\bar{b}$'' trigger, assuming a Higgs to $\pi_v$ branching ratio of 1\%. Only displaced tracks  with IP > 2 mm and $p_T > 1$ GeV are vertexed.
The lightest expected signals require special consideration compared to heavier signals. In order to be sensitive to a $m_{\pi_v} \approx 10$ GeV signal, vertex requirements with low $N_{tracks}$ and $m_{DV}$, \textit{i.e.}  $N_{tracks}\ge 4$ and $m_{DV}\ge 4$ GeV or lower, must be implemented.

We choose vertex selection criteria based on existing tracker searches performed by ATLAS and CMS. The reasoning is two-fold: we want to project the sensitivity of these search techniques for Run II, possibly with different triggers, while remaining confident that there are no more than a few background events. Ideally a background discriminant would be trained for the signal considered here. For a multitrack DV search without requiring any objects associated to the DV or additional event criteria, we choose $N_{track} \geq 5$ and $m_{DV} \geq 10$ for which we expect to be background free for a wide class of triggers as was the case in~\cite{Aad:2015rba}. For a DV associated to a displaced dijet, we choose $N_{track} \geq 4$, $m_{DV} \geq 4$, and $p_T \geq 8$ GeV for our vertex criteria, which is comparable to the vertex requirements of the displaced dijet search~\cite{CMS:2014wda} which we note did not remove regions of detector material and so could possibly be improved. For searches requiring two DVs reconstructed in the tracker, it is likely not necessary to impose a $m_{DV}$ requirement. We require two DVs with $N_{tracks} \geq 5$ for such a search.

We consider five  main search strategies:
\begin{enumerate}[I.]
\item Search for at least one high mass ($m_{DV} > 10 \text{ GeV}$) and high track multiplicity ($N_{tracks} \geq 5$) DV, based on the zero background search done by ATLAS~\cite{Aad:2015rba}.
\item Search for one high track multiplicity ($N_{tracks} \geq 5$) DV together with reconstructing the Higgs boson mass and the two $\pi_v$ particles in the displaced jets of the event. We require 2, 3, or 4 displaced jets, defined as a jet ($p_T > 20$ GeV) with at most 1 prompt track with IP < 0.5 mm and at least two displaced tracks. The invariant mass of all the displaced jets are required to be within $20 \text{ GeV}$ of the Higgs mass. Additionally, all combinations of either one or two displaced jets are checked in order to reconstruct the invariant mass of the two $\pi_v$ particles, which must be within $75\%$ of each other.
\item Search for one DV ($m_{DV} >4 \text{ GeV}$, $N_{tracks} \geq 4$, and $p_T \geq 8$) associated to a displaced dijet, similar to the search by CMS~\cite{CMS:2014wda}. The momentum of the dijet is required to be consistent with the DV position in the detector ($| \hat{p}_{dijet} - \hat{r}_{DV} | < 0.15$).
\item Search for one DV ($m_{DV} >4 \text{ GeV}$, $N_{tracks} \geq 4$, and $p_T \geq 8$) associated to a displaced jet with 2-prong substructure. 
We use the Cambridge-Aachen algorithm with $\Delta R = 0.5$ for this search and tag jets with substructure using the mass drop tagger \cite{Butterworth:2008sd} with
$\mu = 0.67$ and $y = 0.15$ by reversing the jet clustering by one step and requiring a significant drop in invariant mass of each subjet. The two subjets, separated by $\Delta R$, are filtered by reclustering only the constituents of each subjet using a size $\Delta R / 2$. The two filtered subjets must form a dijet whose momentum is consistent with the DV position in the detector ($| \hat{p}_{dijet} - \hat{r}_{DV} | < 0.15$).
\item Search for 2 DVs ($N_{tracks} \geq 5$) in the same event, comparable to the ATLAS search~\cite{Aad:2015uaa}.
\end{enumerate}

These strategies could be implemented separately or combined in order to optimally reduce background. Searches I, III, and IV achieve sensitivity to signals with final states involving only a single DV per event.

\subsection{Results\label{sec:Results}}
\begin{figure*}[tp!]
 \vspace{1cm}   
\subfloat[\label{searchI}]{\includegraphics[width=0.48\textwidth]{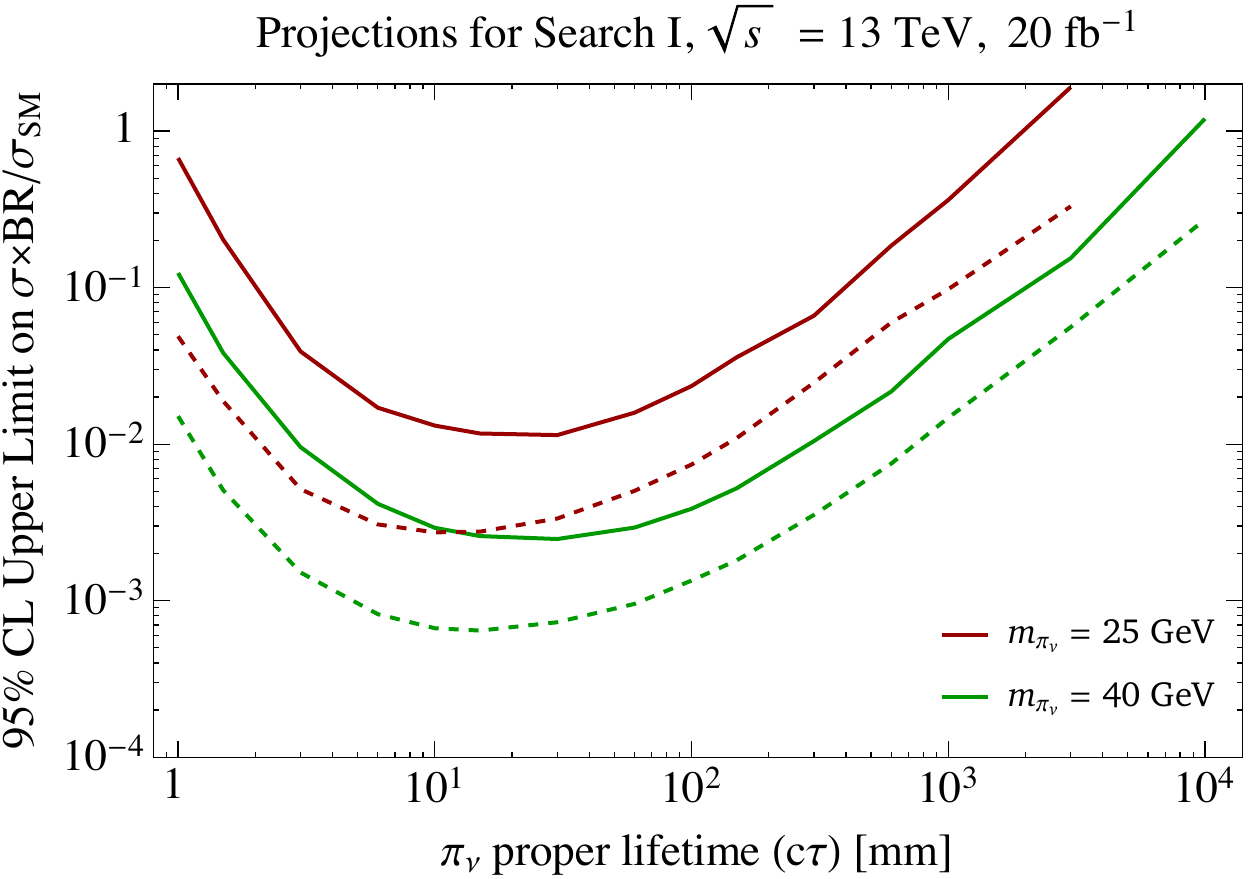}}
\hfill
\subfloat[\label{searchII}]{\includegraphics[width=0.48\textwidth]{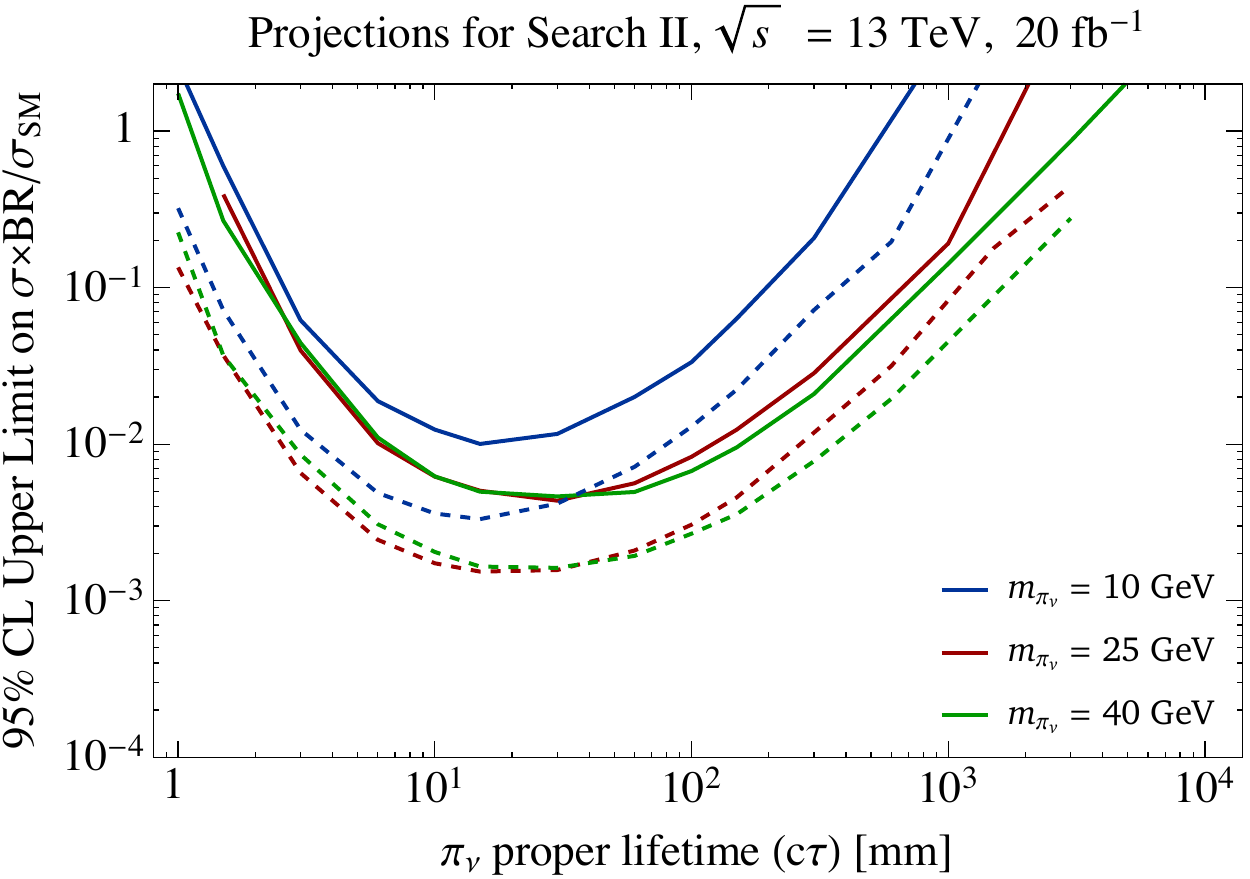}}\\
\subfloat[\label{searchIII}]{\includegraphics[width=0.48\textwidth]{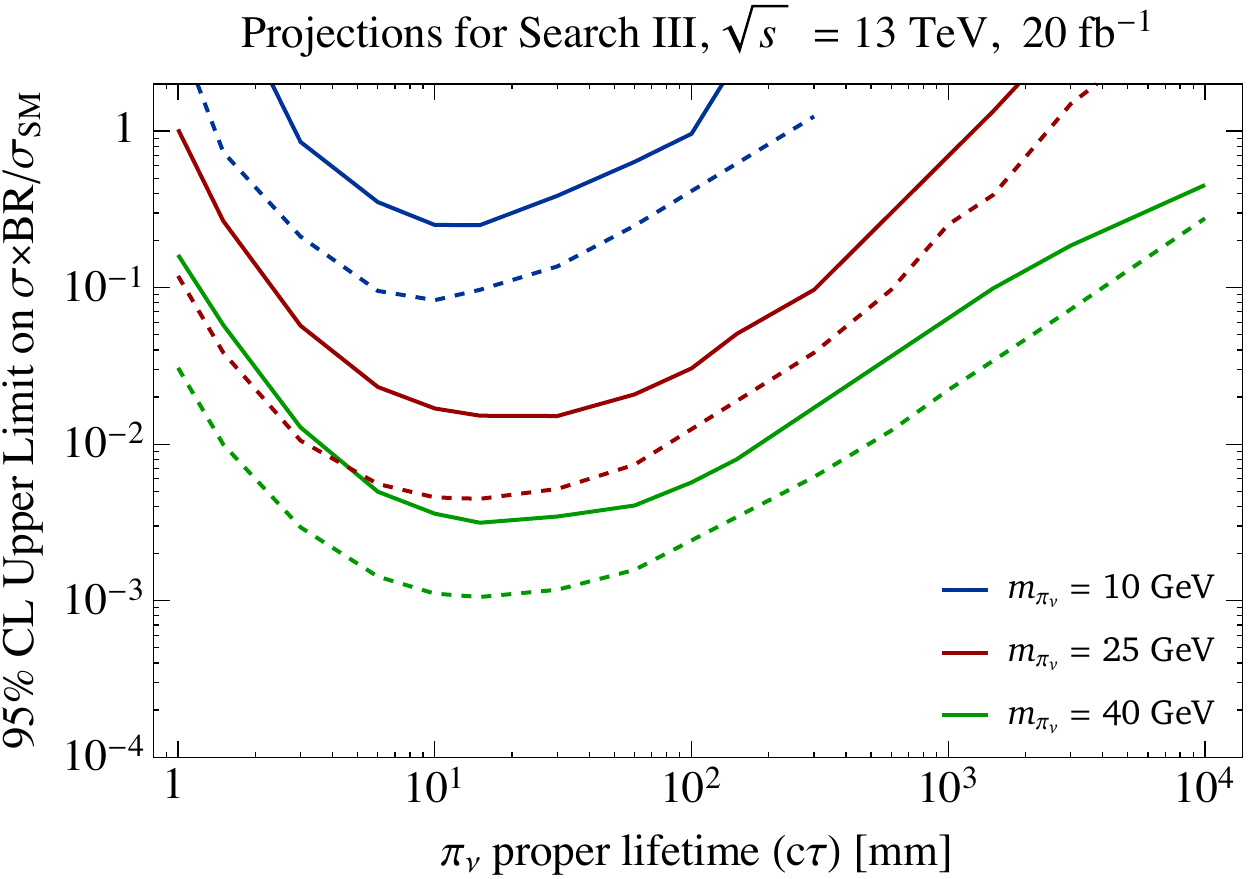}}
\hfill
\subfloat[\label{searchIV}]{\includegraphics[width=0.48\textwidth]{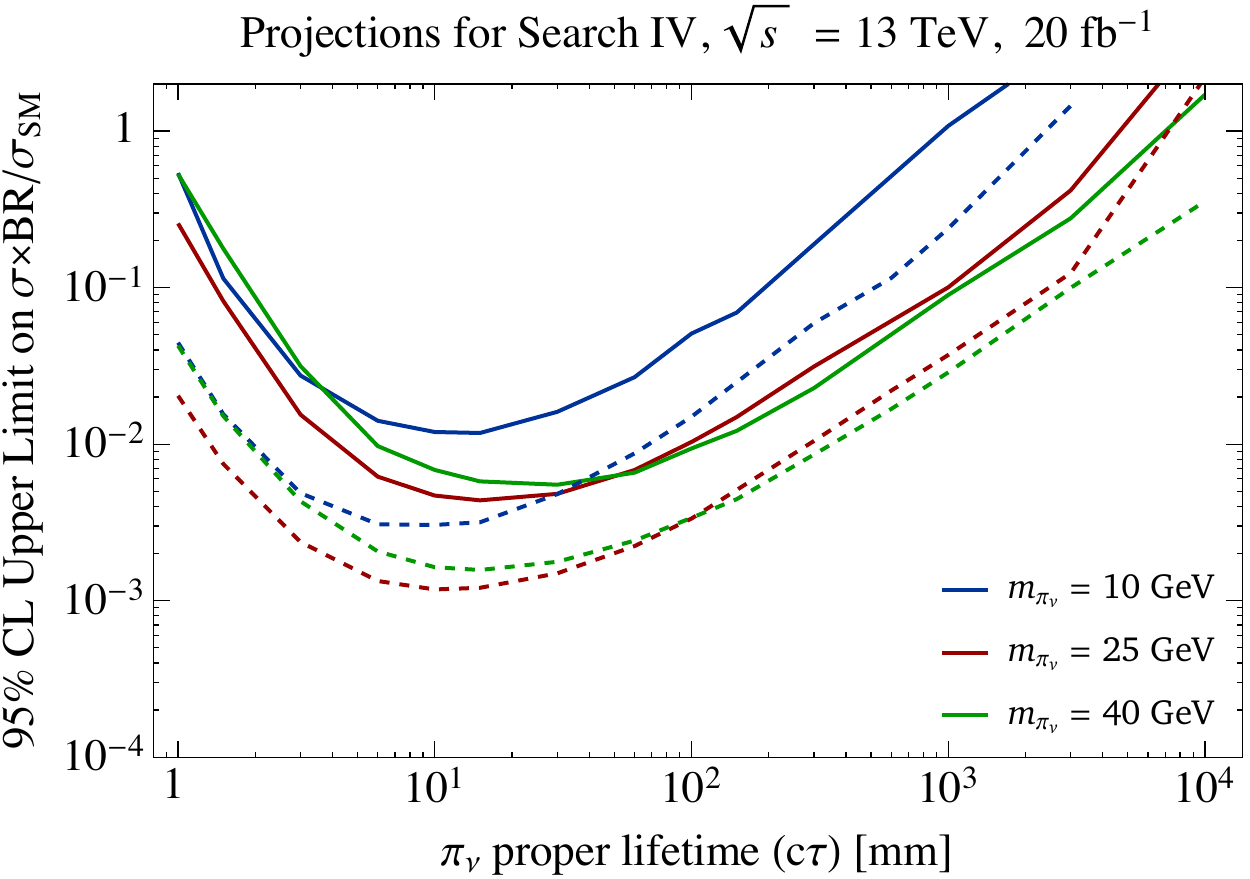}}\\ 
\subfloat[\label{searchV}]{\includegraphics[width=0.48\textwidth]{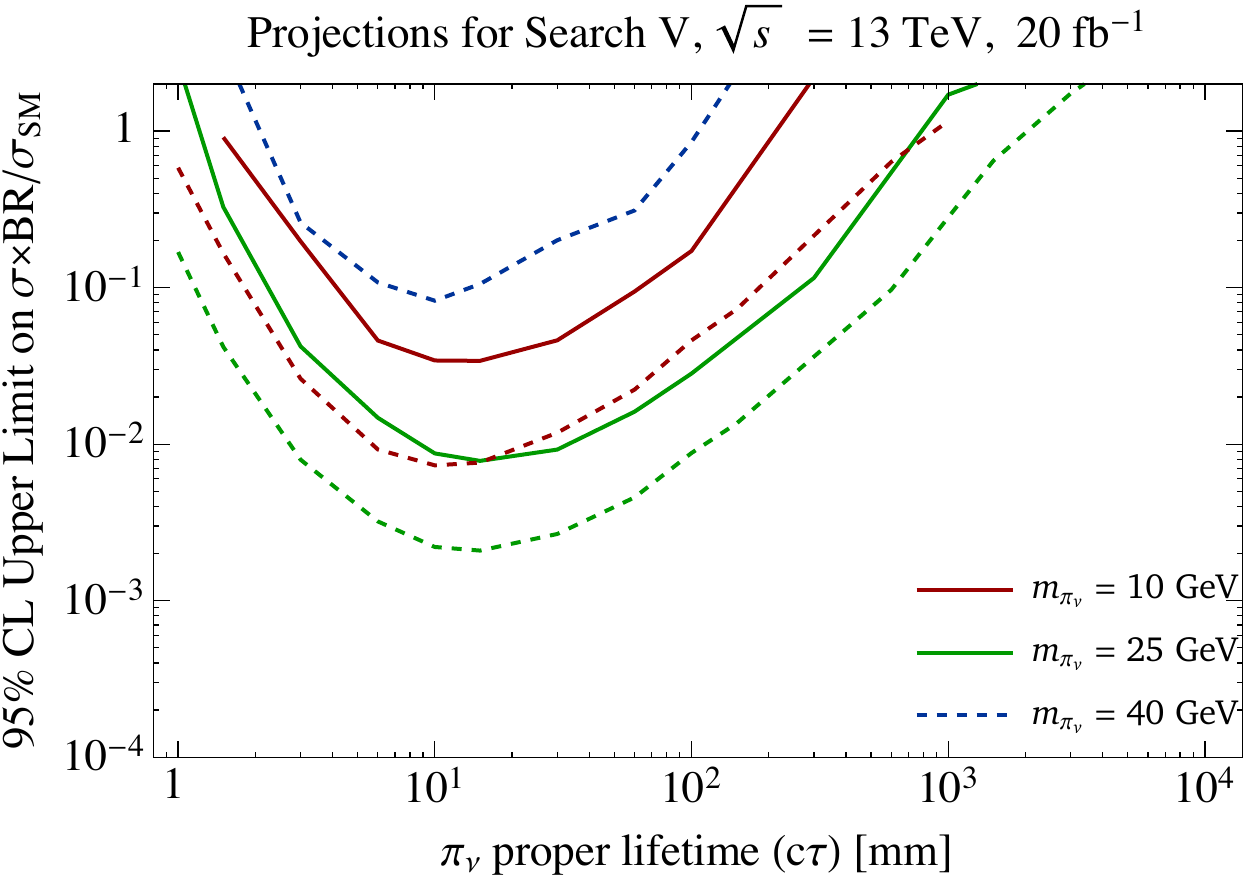}}
\hfill
\subfloat[\label{searchcomb}]{\includegraphics[width=0.48\textwidth]{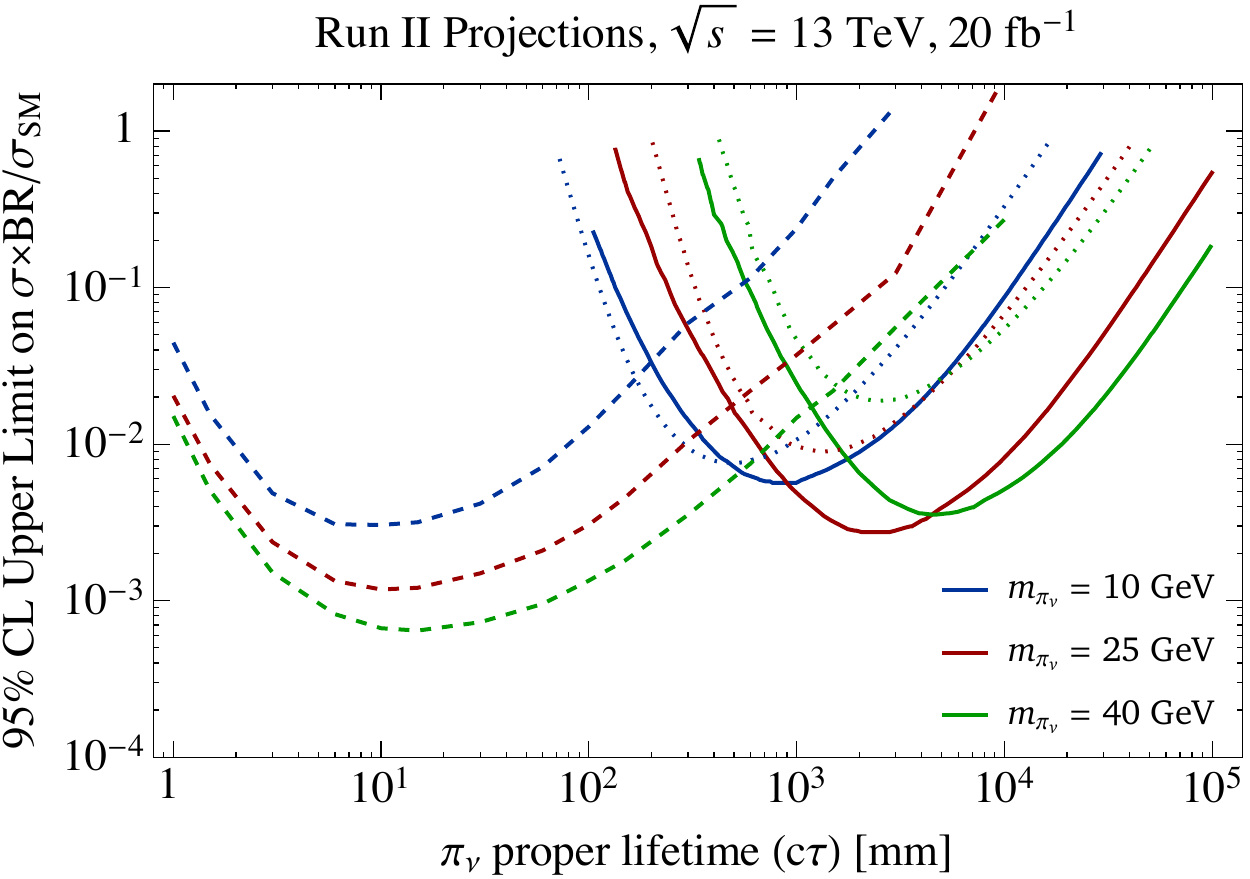}}      
 \caption{Projected sensitivity for Search I (a), Search II (b), Search III (c), Search IV (d) and Search V (e)  with the the ``VBF, $h \rightarrow bb$'' trigger (dashed) and the ``displaced jet'' trigger (solid). Figure (f) shows the overall best sensitivity of all five proposed tracker searches (dashed) together with projected ATLAS search results, rescaled for $\sqrt{s} = 13$~TeV Higgs production.}
 \vspace{2cm}   
 \end{figure*}

We present the projected sensitivities of the search strategies considered in this paper for $\sqrt{s} = 13$ TeV assuming a single background event. It is possible that more background events may pass the trigger and vertex requirements, so this assumption may not be valid. We present results using two triggers, the ``displaced jet'' and ``VBF, $h \rightarrow b\bar{b}$''  triggers, as these have the largest signal efficiency and give sensitivities to the full range of lifetimes relevant to tracker searches. For search strategies with nonzero backgrounds, other triggers in Table \ref{tab:trigger_requirements} may give stronger bounds. We use the CMS detector geometry and displaced tracking efficiency for projecting sensitivities, although the results for the ATLAS detector are not qualitatively different based on our fast detector simulation. All searches lose sensitivity approximately linearly with $c\tau$ for long lifetimes according to the fraction of decays which occur in the inner detector, and exponentially with $c\tau$ for short lifetimes due to tracks failing to pass the displaced IP requirements.

In Fig.~\ref{searchI} we show the results for Search I, a single DV in the event with requiring additional associated objects. While this strategy has strong sensitivity to heavy signals, the sensitivity depends strongly on $m_{\pi_v}$ due to the $m_{DV} > 10$ GeV requirement and has no sensitivity to the $m_{\pi_v} = 10$ GeV signal.

In Fig.~\ref{searchII} we present sensitivities for Search II, which is model-specific to displaced Higgs decays and requires both $\pi_v$ particles to be reconstructed in the displaced jets of the event. By not imposing a $m_{DV}$ cut, the search has sensitivity the full region of expected signal masses.

In Fig.~\ref{searchIII} and~\ref{searchIV}, we show searches requiring a DV associated with a displaced dijet or single displaced jet with 2-prong jet substructure. The two searches are complementary. The separation between the two jets from the $\pi_v$ decay scales as $\Delta R \approx {2 m_{\pi_v}}/{ p_T }$, where $p_T$ is the transverse momentum of the $\pi_v$ particle. The displaced dijet strategy is sensitive for $m_{\pi_v} \gtrsim 20$ GeV at which point the 2 displaced jets from the $\pi_v$ decay typically become merged into a single $\Delta R = 0.5$ jet resulting in stronger bounds from the single displaced jet search.

In Fig.~\ref{searchV} the projected sensitivities for Search V are presented. This search strategy is weaker than others as a result of the weakened tracking efficiency for displaced tracks combined with the requirement of two high track multiplicity vertices in the same event. An improved displaced tracking efficiency, perhaps from future detector upgrades, would drastically improve the efficiency of this search.

%In general, the sensitivity obtained from the ``VBF, $h \rightarrow b\bar{b}$''  trigger is comparable to the dedicated ``displaced jet'' trigger, each peaking at different lifetimes, even though the latter has a higher trigger efficiency. The reason is that the displaced trigger preferentially selects decays with longer lifetimes occurring parts of the detector with weaker tracking efficiency thus reducing vertex reconstruction for events passing the trigger.

The overall sensitivity for a given lifetime for the five proposed searches is shown in Fig.~\ref{searchcomb} together with projected ATLAS search results, obtained by rescaling the 8 TeV limits by the increase in production cross section at $\sqrt{s} = 13$ TeV. Here, it is clear that new tracker searches may have more sensitivity than existing search strategies for signals with lifetimes below 1 m.

\begin{figure}[t!]
\center
\includegraphics[width=.47\textwidth]{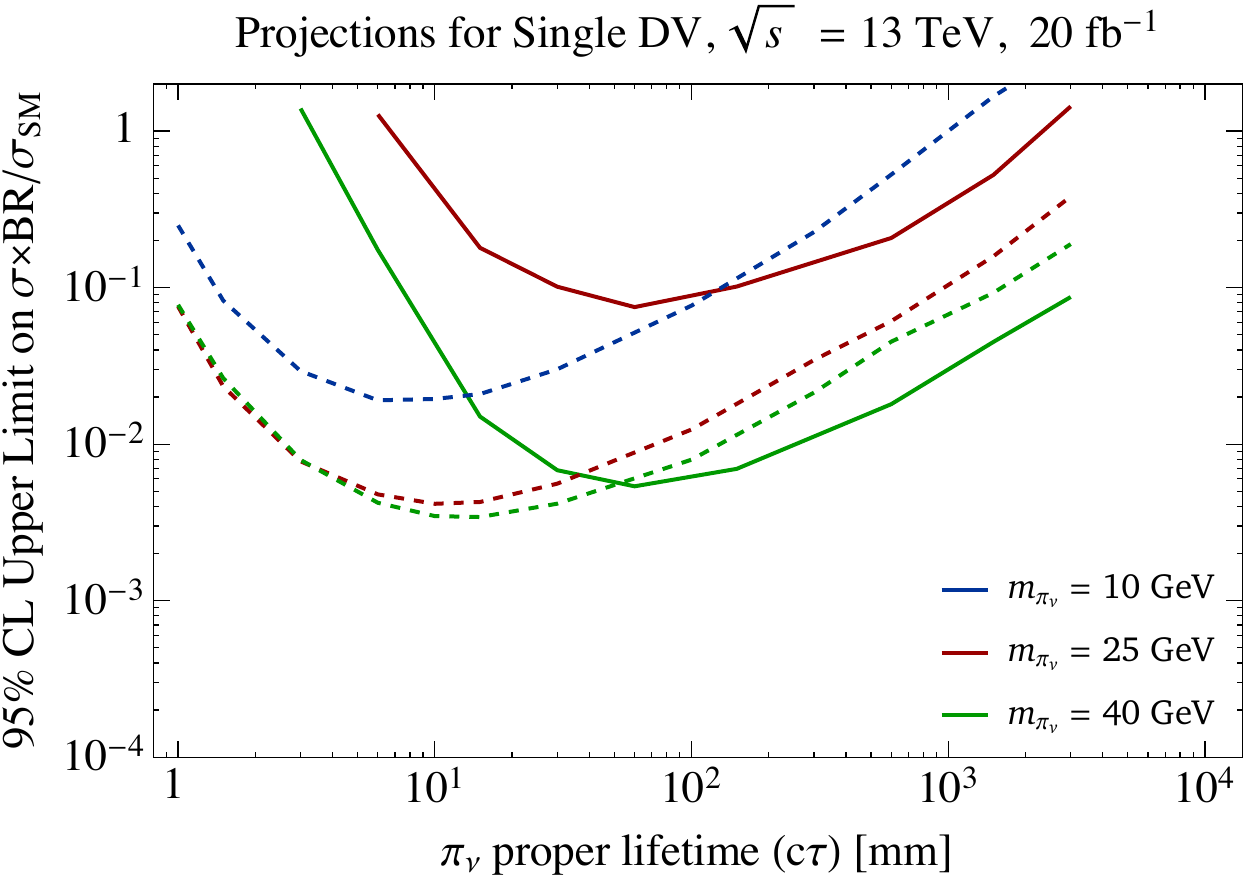}
\caption{Projected sensitivity of Search III with the ``displaced jet'' trigger (solid)  and Search IV with the ``VBF, $h \rightarrow bb$'' trigger (dashed)  for the signal where only one $\pi_v$ is metastable, the other is stable and escapes the detector, contributing to MET. } \label{MET_scenario}
\end{figure}

Finally, we consider a signal where one $\pi_v$ is stable and escapes the detector invisibly resulting in events with only one metastable $\pi_v$ particle. The projected sensitivity for Search III and Search IV are presented for this signal in Fig.~\ref{MET_scenario}. In order to obtain the best sensitivity, Search III was applied to the $m_{\pi_v} = 25$ and 40 GeV signals and Search IV for the $m_{\pi_v} = 10$ GeV signal.

\section{Summary}
\begin{figure}[t!]
\includegraphics[width=.48\textwidth]{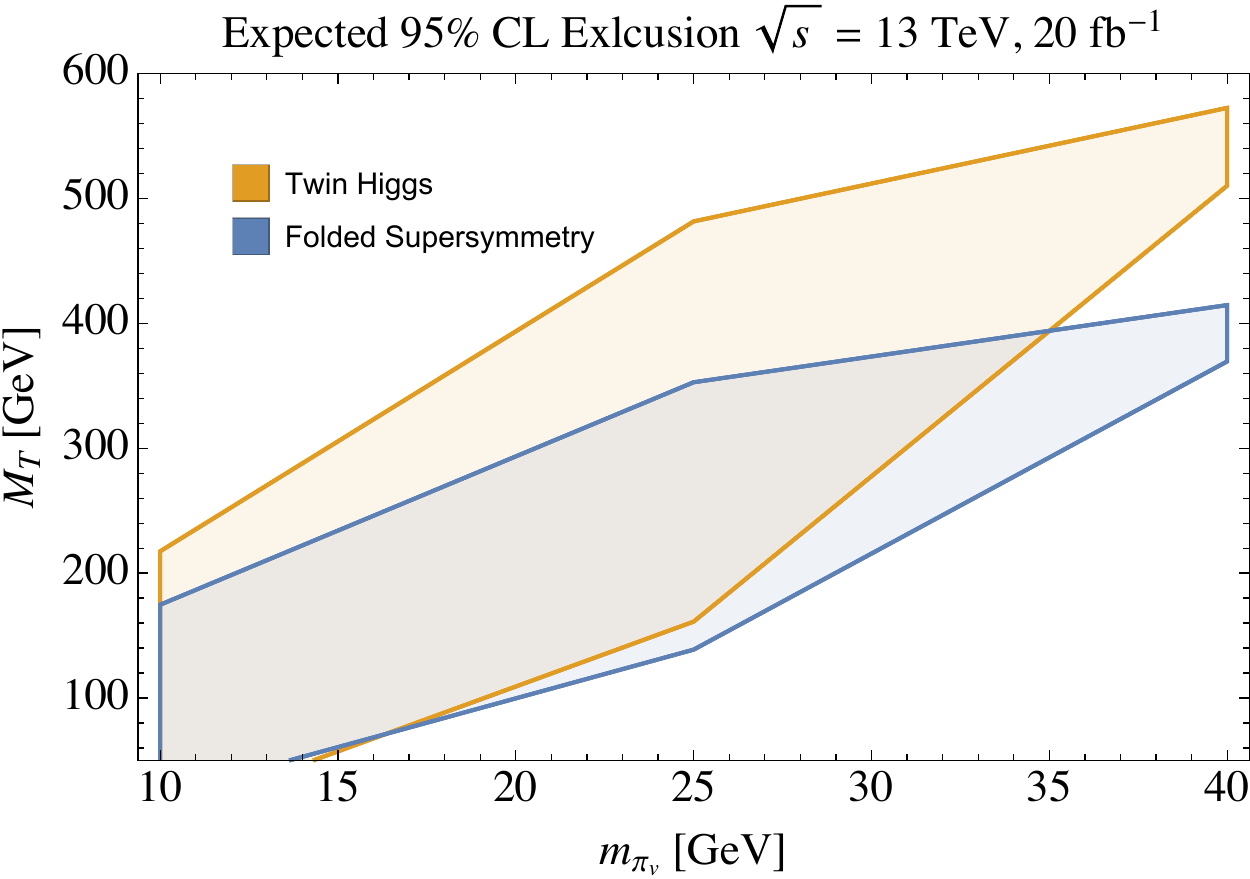}
\caption{Projected 95\% CL excluded region in the $m_{\pi_v}-M_T$ parameter space from the combined Run II search sensitivities show in Fig.~\ref{searchcomb} for the Twin Higgs and Folded Supersymmetry scenarios. See \cite{Curtin:2015fna} for relevant relationships between the masses, glueball lifetimes, and Higgs branching ratios.} \label{toppartner}
\end{figure}

In this study, we have presented existing Run I bounds on displaced Higgs decays and have projected sensitivities for new Run II tracker searches in order to achieve sensitivity to signal lifetimes below 100 mm, which are only weakly constrained by existing searches. Light signals with $m_{\pi_v} \lesssim 10$ GeV will be an experimental challenge to detect above detector backgrounds but can be probed by utilizing new DV search techniques, such as reconstructing the Higgs boson and $\pi_v$ particles in a displaced jet search (Search II) or by searching for a displaced jet associated with jet substructure (Search IV).
The overall sensitivity of these searches, representing the best possible bound for a given mass and lifetime, is shown in Fig.~\ref{searchcomb}. Also shown are the projected ATLAS searches from Run I, rescaled to $\sqrt{s}=13$ TeV. 

We find that Run II searches for DVs with 20 fb$^{-1}$  can probe BRs below 0.1\%, the rates expected by naturalness in the Fraternal Twin Higgs model. In Fig.~\ref{toppartner}, the excluded parameter space from Fig.~\ref{searchcomb} has been translated into a model-dependent excluded region as a function of $m_{\pi_v}$ and the the top partner mass, $M_T$, for the Fraternal Twin Higgs and Folded Supersymmetry models with mirror glueballs (corresponding to the $\pi_v$ particles). The optimistic value of mirror sector hadronization parameter $\kappa=1$, as defined in ~\cite{Curtin:2015fna}, was used.   We find that the Run II LHC can probe twin tops as heavy as 575 GeV for the Fraternal Twin Higgs and stops as heavy as 400 GeV in Folded SUSY.

\section{Acknowledgements}
We thank Yanou Cui, Yuri Gershtein, Eva Halkiadakis, Yonit Hochberg, Nimrod Taiblum, Noam Tal Hod, Matthew Walker, and Margaret Zientek for useful discussions.  We are grateful to David Curtin and Abner Sofer for comments on the manuscript. We thank David Curtin for providing his code for translating our  exclusion curves to model parameters. We thank David Curtin and Yanou Cui for help interpreting the CMS displaced jet trigger. C.C., E.K. and S.L. are supported in part by the NSF grant PHY-1316222. EK is supported by a Hans Bethe Postdoctoral Fellowship at Cornell. O.S. is supported in part by a grant from the Israel Science Foundation.

\bibliographystyle{h-physrev}
\bibliography{Arxiv}

\begin{thebibliography}{10}

\bibitem{Chacko:2005pe}
Z.~Chacko, H.-S. Goh, and R.~Harnik,
\newblock Phys.Rev.Lett. {\bf 96}, 231802 (2006), hep-ph/0506256.

\bibitem{Barbieri:2005ri}
R.~Barbieri, T.~Gregoire, and L.~J. Hall,
\newblock (2005), hep-ph/0509242.

\bibitem{Chacko:2005vw}
Z.~Chacko, Y.~Nomura, M.~Papucci, and G.~Perez,
\newblock JHEP {\bf 0601}, 126 (2006), hep-ph/0510273.

\bibitem{foldedsusy}
G.~Burdman, Z.~Chacko, H.-S. Goh, and R.~Harnik,
\newblock JHEP {\bf 0702}, 009 (2007), hep-ph/0609152.

\bibitem{quirky}
H.~Cai, H.-C. Cheng, and J.~Terning,
\newblock JHEP {\bf 0905}, 045 (2009), 0812.0843.

\bibitem{Craig:2014aea}
N.~Craig, S.~Knapen, and P.~Longhi,
\newblock Phys.Rev.Lett. {\bf 114}, 061803 (2015), 1410.6808.

\bibitem{orbifold}
N.~Craig, S.~Knapen, and P.~Longhi,
\newblock JHEP {\bf 1503}, 106 (2015), 1411.7393.

\bibitem{Carmi:2012in}
D.~Carmi, A.~Falkowski, E.~Kuflik, T.~Volansky, and J.~Zupan,
\newblock JHEP {\bf 1210}, 196 (2012), 1207.1718.

\bibitem{Carmi:2012yp}
D.~Carmi, A.~Falkowski, E.~Kuflik, and T.~Volansky,
\newblock JHEP {\bf 1207}, 136 (2012), 1202.3144.

\bibitem{Burdman:2014zta}
G.~Burdman, Z.~Chacko, R.~Harnik, L.~de~Lima, and C.~B. Verhaaren,
\newblock Phys.Rev. {\bf D91}, 055007 (2015), 1411.3310.

\bibitem{Juknevich:2009gg}
J.~E. Juknevich,
\newblock JHEP {\bf 08}, 121 (2010), 0911.5616.

\bibitem{Strassler:2006ri}
M.~J. Strassler and K.~M. Zurek,
\newblock Phys. Lett. {\bf B661}, 263 (2008), hep-ph/0605193.

\bibitem{fraternal}
N.~Craig, A.~Katz, M.~Strassler, and R.~Sundrum,
\newblock (2015), 1501.05310.

\bibitem{Kang:2008ea}
J.~Kang and M.~A. Luty,
\newblock JHEP {\bf 11}, 065 (2009), 0805.4642.

\bibitem{Juknevich:2009ji}
J.~E. Juknevich, D.~Melnikov, and M.~J. Strassler,
\newblock JHEP {\bf 07}, 055 (2009), 0903.0883.

\bibitem{Curtin:2015fna}
D.~Curtin and C.~B. Verhaaren,
\newblock (2015), 1506.06141.

\bibitem{Aad:2015rba}
ATLAS, G.~Aad {\em et~al.},
\newblock (2015), 1504.05162.

\bibitem{Aad:2015uaa}
ATLAS, G.~Aad {\em et~al.},
\newblock (2015), 1504.03634.

\bibitem{Aad:2015asa}
ATLAS, G.~Aad {\em et~al.},
\newblock Phys.Lett. {\bf B743}, 15 (2015), 1501.04020.

\bibitem{CMS:2014wda}
CMS, V.~Khachatryan {\em et~al.},
\newblock Phys.Rev. {\bf D91}, 012007 (2015), 1411.6530.

\bibitem{Han:2007ae}
T.~Han, Z.~Si, K.~M. Zurek, and M.~J. Strassler,
\newblock JHEP {\bf 07}, 008 (2008), 0712.2041.

\bibitem{Alwall:2011uj}
J.~Alwall, M.~Herquet, F.~Maltoni, O.~Mattelaer, and T.~Stelzer,
\newblock JHEP {\bf 1106}, 128 (2011), 1106.0522.

\bibitem{Sjostrand:2014zea}
T.~Sjostrand {\em et~al.},
\newblock Comput.Phys.Commun. {\bf 191}, 159 (2015), 1410.3012.

\bibitem{deFavereau:2013fsa}
DELPHES 3, J.~de~Favereau {\em et~al.},
\newblock JHEP {\bf 1402}, 057 (2014), 1307.6346.

\bibitem{Cacciari:2011ma}
M.~Cacciari, G.~P. Salam, and G.~Soyez,
\newblock Eur.Phys.J. {\bf C72}, 1896 (2012), 1111.6097.

\bibitem{Csaki:2015uza}
C.~Csaki, E.~Kuflik, S.~Lombardo, O.~Slone, and T.~Volansky,
\newblock (2015), 1505.00784.

\bibitem{Abazov:2009ik}
D0, V.~Abazov {\em et~al.},
\newblock Phys.Rev.Lett. {\bf 103}, 071801 (2009), 0906.1787.

\bibitem{Aad:2013txa}
ATLAS, G.~Aad {\em et~al.},
\newblock JINST {\bf 8}, P07015 (2013), 1305.2284.

\bibitem{Schwaller:2015gea}
P.~Schwaller, D.~Stolarski, and A.~Weiler,
\newblock JHEP {\bf 1505}, 059 (2015), 1502.05409.

\bibitem{ATLAS:2012yna}
ATLAS,
\newblock (2012).

\bibitem{Kaplan:2011vf}
D.~E. Kaplan and M.~McEvoy,
\newblock Phys.Rev. {\bf D83}, 115004 (2011), 1102.0704.

\bibitem{Note1}
See for instance, figures 9-10 in \cite {Aad:2015rba}.

\bibitem{Aaltonen:2011rja}
CDF, T.~Aaltonen {\em et~al.},
\newblock Phys.Rev. {\bf D85}, 012007 (2012), 1109.3136.

\bibitem{Butterworth:2008sd}
J.~M. Butterworth, A.~R. Davison, M.~Rubin, and G.~P. Salam,
\newblock AIP Conf. Proc. {\bf 1078}, 189 (2009), 0809.2530.

\end{thebibliography}
\end{document}